\documentclass[12pt]{article}
\usepackage{a4wide,epsfig}

\begin{document}

\title{High Order QED Corrections in Physics of Positronium\footnote{
Talk given at  the Workshop on Positronium Physics, ETH Hönggerberg,
Zurich, May 30-31, 2003.}}

\author{Alexander A. Penin$^{a,b}$\\[5mm]
{\small\it
$^{a}$II. Institut f\"ur Theoretische Physik, Universit\"at Hamburg,}\\
{\small\it 
Luruper Chaussee 149, 22761 Hamburg, Germany}\\
{\small\it 
$^{b}$Institute for Nuclear Research, Russian Academy of Sciences,}\\
{\small\it 
60th October Anniversary Prospect 7a, 117312 Moscow, Russia}
}

\date{}
\maketitle

\begin{abstract}
High-order perturbative corrections  to
positronium decays and hyperfine splitting 
are briefly  reviewed. Theoretical
predictions are compared to  the most recent experimental data.
Perspectives of future calculations are discussed. 
\end{abstract}

\section{Introduction}

Positronium,  an electromagnetic bound state of the electron $e^-$ and
the positron $e^+$, is the lightest known atom.
Thanks to the smallness of the electron mass $m_e$ the strong and
weak interaction effects are negligible and its properties can be calculated
perturbatively in quantum electrodynamics (QED) as an expansion in
Sommerfeld's fine structure constant $\alpha$ with very high precision only
limited by the complexity of the calculations.
Positronium is thus a unique laboratory for testing the QED theory of weakly
bound systems.   The theoretical analysis is, however, complicated 
in comparison to other hydrogen-like atoms by
annihilation and recoil effects.
At the same time due to  negligible 
short-distance effects of  the  virtual strongly  
interacting    heavy  particles,
positronium could be a sensitive probe
of the ``new physics'' at long distance such as
large extra dimensions, mirror universe,
hypothetical super-weakly interacting massless particles, {\it etc.}.  
Naturally, positronium is  a subject of   extensive
theoretical and experimental investigations.

Besides its phenomenological importance the positronium system is also
very interesting from the theoretical point of view because it possesses a
highly sophisticated multiscale dynamics and its study demands the full power
of the effective field theory approach.

In this paper we give a brief review of the current status
and future perspectives 
of the high order perturbative analysis of positronium in QED.
We focus on the positronium ground state 
decay rates and hyperfine splitting (HFS)
as the most promising observables for precise experimental measurements.

\section{Positronium decays}
The present theoretical knowledge of the 
decay rates (widths) of the $^3S_1$ orthoposit\-ronium (o-Ps)
and $^1S_0$ parapositronium (p-Ps) 
ground states 
to two and three photons, respectively, may be summarized as follows
\begin{eqnarray}
\Gamma_o^{\rm th}&=&{2(\pi^2-9)\alpha^6m_e\over 9\pi}
\left\{1+{\alpha\over\pi}10.286\,606 (10)+
\left({\alpha\over\pi}\right)^2\left[
{\pi^2\over 3}\ln{\alpha}
+44.87(26)\right]\right.
\nonumber\\
&&\left.
+{\alpha^3\over\pi}\left[-{3\over2}\ln^2{\alpha} 
+\left(3.428\,869(3)-{229\over 30}+8\ln2\right)\ln{\alpha}
+{D_o\over \pi^2}\right]\right\}\,,
\label{sort}
\\
\Gamma_p^{\rm th}&=&{\alpha^5m_e\over 2}
\left\{1+{\alpha\over\pi}\left({\pi^2\over 4}-5\right)
+
\left({\alpha\over\pi}\right)^2\left[
-2\pi^2\ln{\alpha}
+5.1243(33)\right]\right.
\nonumber\\
&&\left.
+{\alpha^3\over\pi}\left[-{3\over2}\ln^2{\alpha} 
+\left({533\over 90}-{\pi^2\over 2}+10\ln{2}\right)\ln{\alpha}
+{D_p\over \pi^2}\right]\right\}\,.
\label{spar}
\end{eqnarray}
The first order corrections to  o-Ps and p-Ps decay rates
have been found in Refs.~\cite{CLS} and \cite{HarBra}. 
The logarithmically enhanced second order corrections 
have been obtained in Refs.~\cite{CasLep1} and \cite{KhrYel}.
Recently, the calculation of the 
nonlogarithmic  second order corrections 
has been completed  in Refs.~\cite{AFS1} and \cite{CMY2}. 
In the third order only the double  logarithmic \cite{Kar} and 
single  logarithmic corrections \cite{KniPen1} are known.\footnote{
In Eq.~(\ref{sort}) the nonanalytic
part of the $\alpha^3\ln{\alpha}$ coefficient represents
the interference  between the first order correction  and the second
order logarithmic term.}
The coefficients $D_{o,p}$ parameterize the unknown nonlogarithmic
${\cal O}(\alpha^3)$ terms.  
The o-Ps decays into  five  photons and   p-Ps decays into  four photons, 
which are not included in 
Eqs.~(\ref{sort},~\ref{spar}), lead to an increase of the numerical
coefficients in front of $(\alpha/\pi)^2$ by  
$0.187(11)$ and $0.274290(8)$, respectively \cite{AdkBro}.
Including all the terms known so far  we obtain for the p-Ps and o-Ps total
decay rates
\begin{eqnarray}
\Gamma_o^{\rm th}&=&7.039979(11)~\mu{\rm s}^{-1}\,,
\label{thort}
\\
\Gamma_p^{\rm th}&=&7989.6178(2)~\mu{\rm s}^{-1}\,,
\label{thpar}
\end{eqnarray}
where the given errors stem only from the uncertainty in the numerical
values of the perturbative coefficients  
and we postpone  the discussion of the total
uncertainty of theoretical estimates to   Sec.~\ref{therr}.

Experimental study of the o-Ps decay rate has a colorful history of
inconsistent results and poor agreement with  theoretical predictions
(see Ref.~\cite{AFS2} for a review).
The most recent independent measurements in $SiO_2$ powder 
\cite{Tok} and vacuum \cite{Anno} experiments
\begin{eqnarray}
\Gamma_o^{\rm exp}
&=&7.0396(12)^{\rm stat.}(11)^{\rm syst.}~\mu{\rm s}^{-1}
~~{\rm (SiO_2~ powder)}\,,
\label{exportp}\\
\Gamma_o^{\rm exp}&=&7.0404(10)^{\rm stat.}(8)^{\rm syst.}
~\mu{\rm s}^{-1}~~{\rm (vacuum)}\,,
\label{exportv}
\end{eqnarray}
however, are in a very good agreement with Eq.~(\ref{thort}).
The most recent experimental data on p-Ps decay rate \cite{Annp}  
\begin{eqnarray}
\Gamma_p^{\rm exp}&=&7990.9(1.7)~\mu{\rm s}^{-1}
\label{exppar}
\end{eqnarray}
is also consistent with Eq.~(\ref{thpar}) within the error bars.

\section{Positronium  HFS}
Positronium HFS,
$\Delta\nu=E\left(1^3S_1\right)-E\left(1^1S_0\right)$,
where $E\left(1^1S_0\right)$ and $E\left(1^3S_1\right)$ are the energy levels
of p-Ps and o-Ps ground state,  is the most precisely measured
quantity in positronium spectroscopy as far as the absolute precision is
concerned. The most recent measurements of  HFS  yielded \cite{Mil,Rit}
\begin{eqnarray}
\Delta\nu^{\rm exp}&=&203.387\,5(16)\,\mbox{GHz}\,,  
\label{exphfs1}
\\
\Delta\nu^{\rm exp}&=&203.389\,10(74)\,\mbox{GHz}\,. 
\label{exphfs2}
\end{eqnarray}
On the  theoretical side we have
\begin{eqnarray}
\Delta\nu^{\rm th}&=&{7m_e\alpha^4\over 12}\left\{1
-\frac{\alpha}{\pi}\left(\frac{32}{21}+\frac{6}{7}\ln2\right)
+\left(\frac{\alpha}{\pi}\right)^2
\left[-\frac{5}{14}\pi^2\ln{\alpha}+\frac{1367}{378}-\frac{5197}{2016}\pi^2
\right.\right.
\nonumber\\
&&
\left.+\left(\frac{6}{7}+\frac{221}{84}\pi^2\right)\ln2
-\frac{159}{56}\zeta(3)\right]
+\frac{\alpha^3}{\pi}\left[-\frac{3}{2}\ln^2{\alpha}
-\left(\frac{62}{15}-\frac{68}{7}\ln2\right)\ln{\alpha}\right.
\nonumber\\
&&
\left.\left.
+{D_\nu\over \pi^2}
\right]\right\}\,.
\label{shfs}
\end{eqnarray}
The first order correction has been calculated in Ref.~\cite{KarKle}.
The logarithmically enhanced second order correction has been found in
Refs.~\cite{CasLep1,BodYen}.  
The nonlogarithmic second order term includes the contribution due to
the radiative correction to the Breit potential \cite{BroEri}, the three-,
two- and one-photon annihilation contributions \cite{ABZ}, the
nonannihilation radiative recoil contribution \cite{STY}, and the pure recoil
correction computed numerically in Ref.~\cite{Pac1} and analytically in
Ref.~\cite{CMY1}.
In the third order, as in the case of the decay rates,
only the  double logarithmic \cite{Kar} 
and  single logarithmic corrections \cite{Hil} are known.
Collecting all the available contributions we get
\begin{eqnarray}
\Delta\nu^{\rm th}&=&203.391\,69\,\mbox{GHz}\,,
\end{eqnarray}
which exceeds Eqs.~(\ref{exphfs1}) and (\ref{exphfs2}) 
by approximately 2.6 and 3.5
experimental standard deviations, respectively.

\section{Structure of QED series and theoretical \\uncertainty}
\label{therr}
As far as pure QED predictions are concerned,
the accuracy of the current  theoretical estimates
is determined by the  missing nonlogarithmic $O(\alpha^3)$ terms.
To estimate the corresponding uncertainty  
we may speculate about the magnitudes of the coefficients $D_{o,p,\nu}$.
Substituting the numerical values for the perturbative coefficients 
in  Eqs.~(\ref{sort},~\ref{spar},~\ref{shfs}) we get
\begin{eqnarray}
\Gamma_o^{\rm th}&=&\Gamma_o^{\rm LO}
\left(1-3.27\alpha
+0.33\alpha^2\ln{\alpha}
+4.53\alpha^2
-0.48\alpha^3 \ln^2{\alpha}
-1.76\alpha^3 \ln{\alpha}
+{D_o\over \pi^3}\alpha^3\right)\,,
\nonumber\\
&&
\label{snumort}\\
\Gamma_p^{\rm th}&=&\Gamma_p^{\rm LO}
\left(1-0.81\alpha
-2.00\alpha^2\ln{\alpha}
+0.55\alpha^2
-0.48\alpha^3 \ln^2{\alpha}
+2.52\alpha^3 \ln{\alpha}
+{D_p\over \pi^3}\alpha^3\right)\,,
\nonumber\\
&&
\label{snumpar}\\
\Delta\nu&=&\Delta\nu^{\rm LO}
\left(1-0.67\alpha
-0.36\alpha^2\ln{\alpha}
-0.67\alpha^2
-0.48\alpha^3 \ln^2{\alpha}
+0.83\alpha^3 \ln{\alpha}
+{D_\nu\over \pi^3}\alpha^3\right)\,.
\nonumber
\\&&
\label{snumhfs}
\end{eqnarray}
As we observe, the perturbative coefficients
do not systematically increase in magnitude with the order of the expansion
if $\alpha$ (not  $\alpha/\pi$) 
is taken as the  formal expansion parameter. Thus we estimate
the numerical value of  $D_{o,p,\nu}/\pi^3$  
to be a few units, which is a typical number in 
Eqs.~(\ref{snumort},~\ref{snumpar},~\ref{snumhfs}).
This naive extrapolation is supported by  the
explicit result for the radiative corrections to HFS in muonium where 
the corresponding coefficient reads \cite{KinNio,NioKin2}
\begin{eqnarray}
{D^{(\mu^+e^-)}_\nu\over \pi^3}&\approx&5\,.
\label{dmuon}
\end{eqnarray}
If the coefficients  $D_{o,p,\nu}$ do not have
absolute 
magnitudes in excess of Eq.~(\ref{dmuon})
then the uncertainties due the lack of their knowledge fall 
within the errors quoted in Table~\ref{tab1}. 
For comparison the minimal reported experimental errors
are also included in this table.\footnote{If given separately, the
statistic and systematic errors are added up in quadrature.}

\begin{table}
  \begin{center}
    \begin{tabular}{|c|c|c|}
 \hline 
&Theory  & Experiment \\ \hline
$\Gamma_o$ &$1.4\cdot 10^{-5}$ $\mu{\rm s}^{-1}$   &
$1.3\cdot 10^{-3}$ $\mu{\rm s}^{-1}$   \\
$\Gamma_p$ &  $1.6\cdot 10^{-2}$$\mu{\rm s}^{-1}$& $1.7 $
$\mu{\rm s}^{-1}$\\
$\Delta\nu$ & 400 kHz & 740  kHz\\ 
    \hline 
    \end{tabular}
    \caption{\label{tab1} Theoretical {\it vs} experimental errors.}
  \end{center}
\end{table}

Further progress in the investigation 
of positronium decays now 
crucially depends  on the reduction of the experimental errors, 
which now greatly exceed the theoretical one.
The experimental error for HFS is compatible with a naive estimate of the
theoretical uncertainty due to as-yet unknown higher order corrections.
Should this discrepancy persist after the dominant terms of the latter have
been calculated and the experimental accuracy has been increased,
this would provide a signal for new physics. 
This makes the HFS to be one of the most interesting topics in positronium
spectroscopy both from the experimental and theoretical points of view.

\section{Theoretical tools and future perspectives}
In  positronium the electron 
and positron are nonrelativistic and have  small relative velocity 
$v\sim \alpha$. 
The dynamics of the  nonrelativistic  bound state is characterized
by three well separated scales: the hard scale of the electron mass
$m_e$, the soft
scale of the bound state momentum  $vm_e$, 
and the ultrasoft scale  of the bound state energy $v^2m_e$.
The presence of several scales and the binding Coulomb
effects essentially complicates 
the high order perturbative calculations which require
a proper conceptual framework. 
The effective  field  theory \cite{CasLep2}  
is now recognized as an ultimate tool
for the analysis of the multiscale systems and, in particular,
for the high order calculations of the nonrelativistic bound 
state parameters. The main idea of the method is to separate
the scales by expanding the QED Lagrangian
in  $v$ and to use  covariant perturbation theory
for the analysis  of hard relativistic modes along with  the
quantum  mechanical description of nonrelativistic soft modes.
In this way the  complicated multiscale problem is decomposed into a 
sequence of simpler problems, each involving  a smaller number of 
scales \cite{CasLep2,NioKin1,PinSot1}. 
The last  advance in the effective theory calculations
is connected to the use of  dimensional 
regularization \cite{CMY1,PinSot2,BSS,KniPen}.
The advantage of this scheme is the absence of  additional 
regulator scales and the simple matching of the 
contributions of different scales. 
Consistent use of  dimensional regularization
in the nonrelativistic effective theory is 
based on the concept of the threshold expansion \cite{BenSmi}.

The effective theory approach is at  the  heart of the 
recent progress in the perturbative QCD bound state 
calculations \cite{KPSS,PenSte}. 
In QED it was successfully applied, in particular,
to the evaluation of  the most complicated 
$O(\alpha^2)$ and $\alpha^3\ln(\alpha)$
terms in Eqs.~(\ref{sort},~\ref{spar},~\ref{shfs})
and forms a solid basis for the future attack on the nonlogarithmic 
third order corrections. 
In the effective theory framework 
the basic ingredients we need to compute  $D_{o,p,\nu}$
are (i) the effective  Hamiltonian 
at ${\cal O}(\alpha v^2)$ and  corresponding
correction to the nonrelativistic Green function,
(ii) the leading retardation effects due to dipole interaction 
of the positronium bound state to the ultrasoft photons, and
(iii) the three-loop hard renormalization 
of the decay or scattering  amplitudes.
In the nonrelativistic part of the calculation
which includes the Hamiltonian, the Green function and the  
retardation effects some results are already available
(see {\it e.g.} Refs.~\cite{Hil,NioKin2,KPSS}) 
and the remaining analysis poses no outstanding technical problem. 
By contrast,  the hard renormalization  constants are given 
by three-loop fully relativistic  on-shell on-threshold
diagrams with four or five 
external lines which are either on the limit or  
beyond  reach of presently available
computational techniques. 
This is  the  main obstacle in the calculation of the 
third order correction.

\section{Summary}
The calculation of missing { ${\cal O}(\alpha^3)$} 
nonlogarithmic terms would be one of the most complicated 
perturbative calculations in quantum field theory though
conceptually the problem is clear, all the necessary
tools are at hand and a number of partial results have been obtained.
Currently the 
experimental uncertainty exceeds the theoretical 
one by two orders of magnitude for positronium decay rates and by a  
factor of two for HFS. Theoretical estimates of the  decay
rates are in perfect agreement with the results of the most recent
experimental measurements while the discrepancy of approximately  $3\sigma$ 
still exists for  HFS. This discrepancy, however, does not look   dramatic 
because in this kind of experiments
the  systematic errors could easily be underestimated.
New measurements of much higher
accuracy  are mandatory to unambiguously 
confirm or confront the   QED predictions
and  to inspire the theorists for the 
${\cal O}(\alpha^3)$  feat.

\section*{Acknowledgments}
The author thanks the organizers of the 
ETH 2003 Workshop on Positronium Physics
for the kind invitation.

\end{document}